# INTEGRATING EMERGING TECHNOLOGIES IN VIRTUAL LEARNING ENVIRONMENTS: A COMPARATIVE STUDY OF PERCEIVED NEEDS AMONG OPEN UNIVERSITIES IN FIVE SOUTHEAST ASIAN COUNTRIES

Roberto Bacani Figueroa Jr., University of the Philippines Open University
Mai Huong Nguyen, Hanoi Open University
Aliza Ali, Open University Malaysia
Lugsamee Nuamthanom Kimura, Sukhothai Thammathirat Open University
Marisa Marisa, Universitas Terbuka
Ami Hibatul Jameel, Universitas Terbuka
Luisa Almeda Gelisan, University of the Philippines Open University

---

## ABSTRACT

*Amid the growing need to keep learners well-informed of the rapid technological advancements brought about by the Fourth Industrial Revolution (4IR), this study investigates the viewpoints of open university students regarding the emerging technology-based virtual learning environments for students at five prominent open universities in Southeast Asia: Hanoi Open University, Open University Malaysia, Sukhothai Thammathirat Open University, University of the Philippines Open University, and Universitas Terbuka. A survey was conducted of undergraduate students to understand their inclinations regarding the features of their virtual learning environments and how well they equip them to be productive citizens and professionals. The results highlight that the students had a significant interest in interactive books and learning analytics. The findings suggest the need to develop a roadmap for open universities to prioritize technological investments and pedagogical strategies to meet the evolving needs of their students in the digital age.*

## INTRODUCTION

The pedagogical landscape of the 21st century is undergoing a revolution as traditional teaching methods, rooted in centuries-old practices, face the realities of a globally connected, digitally driven society. Developments in information and communication technology (ICT) have impacted various aspects of society, specifically how to seek, learn, and share information (Roztocki et al., 2019). ICT impacted the entire educational system by creating a paradigm shift where ITC is a cocreator of knowledge with the student as well as a mentor of and an assessor for the student's progress (Haleem et al., 2022). Bekene Bedada and Machaba (2022) stated that incorporating technology into higher education has become an essential requirement for teaching and learning. As a result, these technologies have also facilitated the emergence of jobs that call for different skills than those required by the workforce of the previous generation.

The need to foster students that are open to critical thinking, able to be creative and adaptive, and have digital literacy is only one of the many new challenges facing educators across the globe. Over time, more interactive and fully engaged approaches that encourage problem-solving, working in teams, and actively engaging with learning are replacing the traditional lecture-based approach. Based on the World Economic Forum (Silva et al., 2022), the skills required to thrive in the Fourth Industrial Revolution (4IR) include complex problem-solving, critical thinking, creativity, computational thinking, collaboration, and cognitive flexibility. These skills are usually collectively referred to as 21st century skills.

Education 4.0 is a trend in the educational sector that has emerged from the 4IR and is characterized by educational development that focuses on skills that make learning more customizable, intelligible, portable, global, and virtual. Simply put, the 4IR can be defined as "the age of intelligence in which continued development of automation and disruptive technologies have changed conventional industrial and manufacturing processes" (Hameed & Hashim, 2022, p. 3).

Education 4.0 is characterized by the use of web and internet technologies, artificial intelligence, machine learning, and other digital technologies (Penprase, 2018). Higher education institutions are requiring the implementation of pedagogical innovations and technological advancements to adapt to evolving educational needs (Chigbu et al., 2023; Díaz-Garcia et al., 2022). Along with this fast innovation of technology comes a new generation of learners characterized by having shorter attention spans, desiring immediate feedback, needing rich visuals, and demanding an active learning process. This inspired our study and led us to question our current learning infrastructure and our way of teaching and learning.

The emergence of creative pedagogical practices in response to these shifting demands has made possible innovative approaches to education and knowledge acquisition. A study by Bekene Bedada and Machaba (2022) exemplifies how these advancements use digital tools and online resources to improve learning outcomes. These tools and resources come from a range of disciplines, such as educational psychology, technology, and cognitive science. According to Alakrash and Abdul Razak (2021), traditional curricula must be adjusted to make them more technology compatible as part of Education 4.0, and they should focus more on students acquiring 21st century skills.

With the help of social networking and open educational resources, more students can access knowledge without needing to meet institutional prerequisites, take a predetermined course, or have a personal instructor. In almost every field, education and training programs need to be established and reformed; thus, there is a need to use and apply new pedagogical and innovative technologies in education (Djumaevich et al., 2019). It is crucial to craft new technologies that will enhance the skill development of students and the general population, which will be used to build advanced knowledge, innovation, and creativity. These emerging technologies must, however, gain acceptance in society, or at the very least among their intended users, just like any other innovation. Thus, the first few steps to this involve understanding students' needs and preferences before implementing innovative technologies to achieve a high adoption rate, especially because the needs of learners vary based on geographical location, culture, and learning modality.

This study aims to address a gap in comprehending the technology requirements and preferences of open university students within the framework of the 4IR. Current research mostly emphasizes the technological integration of emerging technologies in conventional education, neglecting the distinct needs of open university students in specific sociocultural contexts. This study examines students from five prominent open institutions in Southeast Asia, bringing into focus their areas of interest while stressing the necessity for virtual learning environments to correspond with the students' particular settings and problems. The results enhance the existing research by offering region- and country-specific, student-centered, insights. Consequently, it can provide a framework for open and distance elearning (ODeL) institutions in the region to prioritize technical investments and instructional approaches. This guarantees that education stays pertinent, accessible, and able to prepare students for professional opportunities in the swiftly changing digital era.

## REVIEW OF LITERATURE

Several studies have investigated students' perspectives and experiences on the application of ICTs and emerging technologies in education.

### *Student Perceptions on ICT in Education*

In Sweden, a study by Saroia and Gao (2018) was conducted, with 130 university students as respondents, to determine their intentions for using mobile learning management systems. The research model used, which was based on the Technology Acceptance Model, included perceived usefulness, perceived ease of use, academic relevance, university management support, and perceived mobility value as external variables. Among other factors, this study stressed the importance of academic relevance, perceived mobility value, and university administration assistance in building easy-to-use mobile learning management systems.

In 2019, a similar study was conducted at Chongqing Medical University, China, with 450 medical students as respondents on the use of mobile health apps for smartphones (Xu et al., 2022). This study aimed to identify the factors influencing the students' use of the application. Results showed that both perceived usefulness and attitude had a significant effect on the intentions of students to continue using mobile health apps.

Recently, research on the acceptance of using intelligent personal assistants for learning was done in Korea (Choi et al., 2024). The subjects of the study were 1,044 South Korean students who had experience using intelligent personal assistants. The results showed that key factors influencing their acceptance were computer self-efficacy, self-management of learning, trust in the intelligent personal assistants, and a sense of social presence when using such devices.

### *The Fourth Industrial Revolution and Education*

Although the current understanding of the Fourth Industrial Revolution remains limited in many sectors, it has become a widely recognized term and is gaining momentum across various parts of the economy. The 4IR, viewed as a convergence of multiple technologies (Centre for the New Economy and Society, 2018), is now drawing increased interest from policymakers, business professionals, and academics. Although the concept is growing in significance and applicability in different fields, there is no agreement on its definition, and no single explanation fully captures its characteristics, even though it has existed since the early 21st century (Lee et al., 2018). Nevertheless, the advancements from the Second Industrial Revolution, with the invention of combustion engines, and the Third Industrial Revolution, marked by breakthroughs in ICT and automated production, laid the groundwork for the swift spread and integration of the 4IR.

Despite technological advancements, the education sector has been reluctant to embrace technology for teaching and learning, even though robots have been used in education, particularly for teaching STEM subjects, since the 1980s (Tymon, 2013). Additionally, technology use in education has mainly been confined to a didactic approach, where teaching involves personal computers and electronic materials. However, the digital technology driving the 4IR goes beyond computers and ematerials, and it must align with a learner-centered approach to effectively enhance student learning experiences.

Bozalek and Ng'ambi (2015) emphasized that to develop graduates with the desired attributes, higher education institutions (HEIs) must consider technological tools from the students' perspective, emphasizing the importance of user-driven initiatives and cloud-based educational opportunities for learning with advanced technology. Gioiosa and Kinkela (2019) further acknowledged that employers look for individuals who are proficient with technology and possess strong interpersonal skills, indicating that students must feel they are acquiring 4IR skills at HEIs. According to research by Richardson and Shan (2019), students' perceptions of learning have a significant impact on their learning styles, which in turn affects the quality of learning outcomes. Herrador-Alcaide et al. (2019) found that students' perceptions of the online learning environment and their own competence can impact their overall satisfaction with the program and curriculum. More specifically, students in HEIs view technology use as an expected and essential part of the learning process. The availability of study materials and a central location for accessing information or extensive resources related to each module in the course were highlighted as significant benefits (Concannon et al., 2005). Students concurred that technology helped them successfully complete their qualifications and about 41% of students agreed or strongly agreed that they learned from internet activities (Gioiosa & Kinkela, 2019). This is to say that students' perceptions of the 4IR-integrated technologies will undoubtedly assist HEIs in recognizing the role of new technologies in shaping skill requirements and ultimately influencing curriculum development in educational systems.

A study by Voogt and Knezek (2018) found that while some students are well-informed about 4IR technologies and their potential benefits, others lack basic knowledge. This disparity in knowledge can affect how students perceive the relevance of their education in preparing them for future careers. Despite the potential benefits for future careers, students also express concerns about 4IR. Common apprehensions include the fear of job displacement due to automation and the ethical implications of AI (Gull et al., 2023).

To address these concerns and enhance student perceptions of 4IR, HEIs must adapt their curricula to include 4IR competencies. This includes fostering critical thinking, creativity, and digital literacy. Incorporating hands-on experiences with 4IR technologies can also help demystify these tools and demonstrate their practical applications (OECD, 2019). Students who are exposed to practical applications of 4IR technologies in their coursework tend to have more positive attitudes and feel more prepared for the future job market (Acuña et al., 2024, Başgül & Çoştu, 2025). Conversely, those with limited exposure may feel apprehensive or skeptical about the relevance of these technologies.

Education is one of the many industries that the Fourth Industrial Revolution, defined by the convergence of digital, biological, and physical technology, is changing. Understanding how students view 4IR is important because it affects how capable and flexible they are with new technology. To help prepare the future generation for a fast-changing technology context, this paper analyzes student perspectives on 4IR.

While the literature on higher education has reported on perceived student needs regarding technology-based solutions, including those that belong to the 4IR, there is a dearth of reports coming from Southeast Asia and from distance learning institutions. The Southeast Asian region has been home to approximately 1.5 million distance learners. Open and Distance eLearning (ODeL) greatly contributes to the fulfillment of the United Nations Sustainable Development Goals (SDGs), specifically SDG 4 (https://sdgs.un.org/goals/goal4). While institutions in this region struggle to expand the reach of education through ODeL, they are also expected to catch up to ensure that their graduates are equipped with the necessary skills to adapt and thrive in the 4IR era. To address this need, the study investigates the perceived needs of students from five open universities in Southeast Asia. The participating universities are members of the OU5, which is an alliance of five open universities in the Association of Southeast Asian Nations (ASEAN) region.

## PURPOSE OF THE STUDY

The primary objective of the research was to collect student viewpoints regarding the characteristics of the emerging technology-based virtual learning environment for undergraduate students at five prominent open universities in Southeast Asia. To achieve this, we sought to answer three research questions:

1. Which features would learners generally want to see implemented in their learning management system?
2. How did learners' ratings from the five open universities vary in terms of the importance of emerging technological features in a learning management system?
3. What are common student needs in the 4IR that could be addressed by integrating

emerging technologies into their learning management system?

## METHODOLOGY

The study followed a quantitative survey research design with a qualitative follow-up to answer the research questions.

*Research Design*

The research design is illustrated in Figure 1, where round table discussions (RTDs) contributed to the survey design followed by the core activities related to survey collection and quantitative data analysis. To explain the quantitative findings, qualitative data from the open-ended questions were thematically analyzed by us and discussed in another round table discussion.

*Figure 1.*
*Diagrammatic Model of the Research Design*

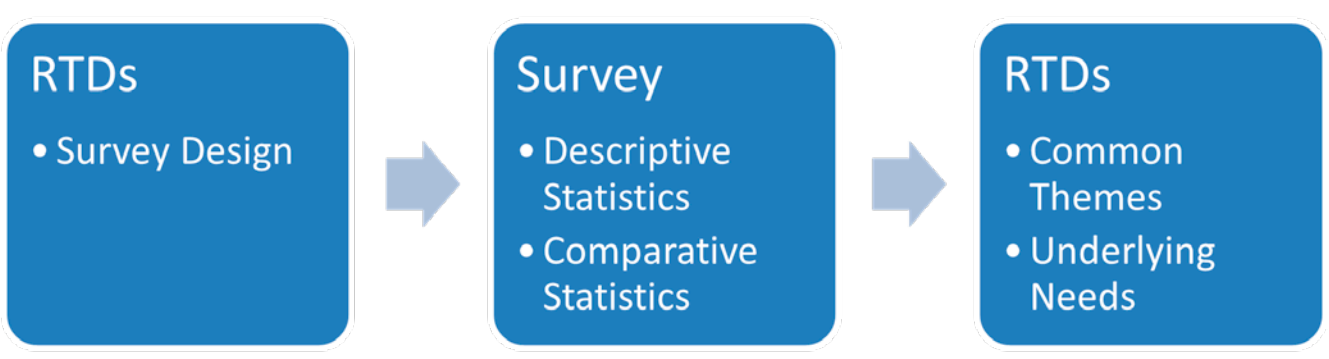

As illustrated in Figure 1, The first set of RTDs explored initial data on students' preferred learning environment, using students' feedback on past course offerings as well as insights from the literature on emerging technologies.

The RTDs were very useful in developing and refining the survey instrument, making it comprehensive and contextually relevant, and allowed us to incorporate items that are significant to the study. It should be noted that the institutions to which we were connected all utilized Moodle, an open-source learning management system (LMS), as the primary platform for delivering online courses.

The survey phase followed, collecting data from 250 participants from the five participating universities. However, once data were collected, another RTD was conducted to discuss both the quantitative and qualitative findings from the survey.

*Survey Instrument*

The initial set of Likert-type items was derived from the literature on emerging educational technologies and student feedback on past course offerings, which were revealed from the first set of RTDs. These items were formulated as statements that students could rate based on their perceived relevance and importance.

The instrument initially had 25 Likert-type items, which were filtered down to 10 items after considering the feasibility of implementing it within the participating universities' virtual learning environments.

The final ten items, each assigned a shorthand label for reporting, were categorized as follows:

1. **AI-assisted feedback:** Provides quick feedback on essays using artificial intelligence (AI).
2. **AI-supported writing/composition:** Assists in essay writing, including plagiarism checks and word suggestions through AI.
3. **Chat-based community:** Enables text-based interactions similar to Discord, Slack, Facebook Messenger, or WhatsApp.
4. **VR simulation/tours:** Facilitates simulation-based learning activities through virtual reality (VR).
5. **VR classes:** Supports real-time interaction with teachers and peers in a 3D virtual classroom.
6. **Metaverse campus:** Provides a fully immersive 3D virtual campus environment.
7. **Interactive videos:** Allows students to take notes or answer embedded quizzes at key moments within videos.
8. **Interactive book:** Offers portable, multimedia-enhanced ebooks that integrate text, audio, video, and self-assessments.
9. **Course suggestions:** Recommends future courses based on students' academic profiles.
10. **Grade prediction:** Predicts student performance based on current scores and course participation.

Since some of the terms used in the items may not be familiar to students, animated graphics interface format (GIF) images accompanied each item to visually illustrate the functionality of each feature. Survey instruments distributed to students in Indonesia, Vietnam, and Thailand were translated in their native languages and validated by a language expert. This approach was also a result of the first set of RTDs. Figure 2 presents an example of one survey item written in English.

*Figure 2.*
*Screenshot of One of the Illustrated, Likert-type Items in the Survey*

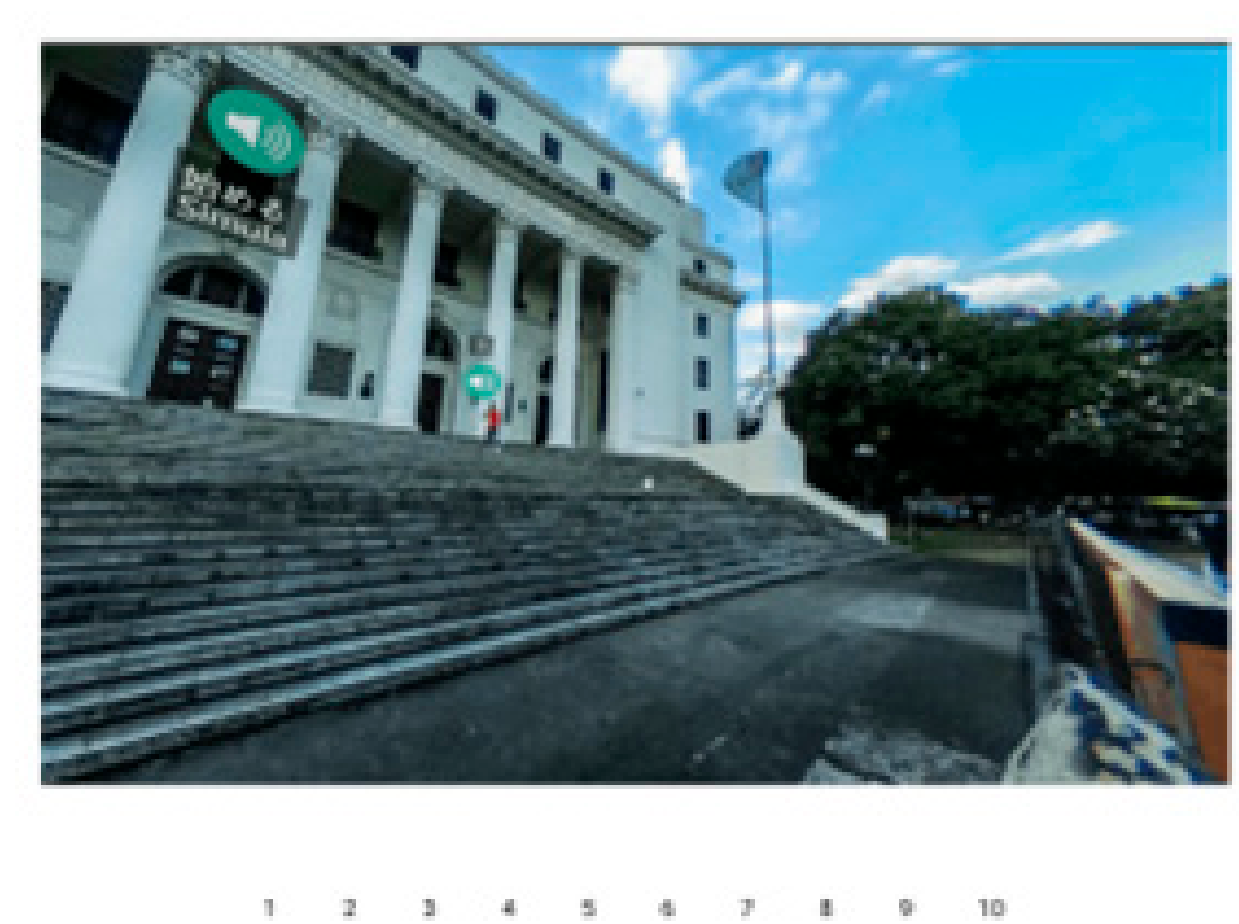

Open-ended questions were included in the instrument to capture student preferences beyond the preidentified features. This ensured that any additional needs or expectations regarding virtual learning environments could be documented and analyzed.

*Ethics*

Prior to data collection, we sought and obtained ethical approval from the Institutional Review Board (IRB) of Mahachulalongkornrajavidyalaya University (MCU) after getting the proposal approved by our respective departments. The submitted research proposal included details on research objectives, methodology, participant recruitment procedures, and data protection protocols, as well as informed consent documentation. The IRB reviewed all supporting documents, such as project outlines, research instruments, and consent forms, to ensure compliance with ethical research standards. Approval was officially granted under IRB approval number (Wor) 012/2566.

*Participants and Context*

The students who participated in this study were from the following five open universities:

- The Hanoi Open University (HOU), Viet Nam, is a renowned university in the field of distance education. It has adapted to the changing landscape of education by creating and implementing a strong virtual learning environment in response to the increasing prevalence of digital platforms. The virtual learning environment of HOU is an extensive digital platform specifically created to enable and enhance distance education. It functions as a virtual platform where students and instructors can engage with one another, like a physical classroom, but with the added advantages of flexibility and accessibility that digital learning offers.
- The Open University Malaysia (OUM), founded in 2000, is a leading institution in distance education. It is known for its pioneering use of a virtual learning environment to provide higher education opportunities to a wide range of learners, including working professionals and lifelong learners. OUM's educational method combines adaptability with technological progress, providing a distinctive learning encounter.
- The Sukhothai Thammathirat Open University (STOU) is a pioneering university in the realm of remote learning. Founded in 1978 as Thailand's inaugural open university, STOU has a notable track record of offering adaptable educational prospects, particularly to individuals who may lack the means to pursue conventional on-campus instruction. The LMS is a vital element of STOU's educational approach, serving as a pivotal tool in its online learning programs.
- The University of the Philippines Open University (UPOU) is a prominent institution in the Philippines that provides degree programs, continuing education courses, and massive open online courses through open and distance elearning. It is one of the constituent units of the University of the Philippines System, the national university of the country. The UPOU, founded on February 23, 1995, was established to provide Filipinos in the country and those based abroad access to quality higher education through open and distance learning. The university was also tasked, through Philippine Republic Act 10650 (Open Distance Learning Law), to assist relevant national agencies, higher education institutions, and technical and vocational institutions in the Philippines to develop their respective distance education programs through capacity building, technical assistance, and research.

- The Universitas Terbuka (UT), located in Indonesia, is a prestigious open institution that has transformed the accessibility of higher education in the country. Founded in 1984, its primary objective is to offer a diverse group of learners the chance to pursue higher education, particularly individuals who face limitations such as job, geographical restrictions, or other obligations that hinder their attendance in conventional on-campus programs.

*Data Collection*

This study employed a volunteer sampling method to select participants. The target population consisted of undergraduate students enrolled in the five open universities. Data from 250 participants, 50 from each university, were collected. The questionnaire was distributed online using Google Forms, allowing easy access for students studying in an open and distance learning context.

*Data Analysis*

The methodologies used in analyzing the data were based on the study's research questions. The first research question was regarding students' most preferred features across the five universities. The data were analyzed using the measures of central tendency, which were computed using the R statistical package. The importance ratings assigned by students were ranked to identify the most highly rated features.

For the second research question, variations in student importance ratings among the five open universities, statistical analyses, including measures of central tendency, ANOVA, and Tukey's Honest Significant Difference (HSD) test, were examined using the R statistical package. Additionally, mean comparisons were visualized through the ggplot2 (https://ggplot2.tidyverse.org/) and multcompView packages (Graves et al., 2025).

The third research question aimed to identify the underlying needs that the preferred features would address. To do this, we participated in a roundtable discussion where we synthesized key themes from our individual reflexive thematic analysis results. In a design-thinking activity, these themes were transformed into keywords and organized using the Mural online platform to identify commonalities across universities.

## FINDINGS

This survey study aimed to identify the current needs of the students that could be addressed by emerging technologies integrated with the learning management systems of five open universities in Southeast Asia. The findings have been organized according to the three research questions of the study.

*Most Needed Features*

The first question was, "Which features would learners generally want to see implemented in their learning management system?" As illustrated in Figure 3, the results showed that the 250 students deemed the interactive book feature as the most important, with a mean of 8.92 out of a maximum of 10. This was followed by features on learning analytics, more specifically, the grade prediction (8.65) and the course suggestion feature (8.59) were generally the second and third highest means, respectively.

*Figure 3.*
*Average Ratings of Desired LMS Features by Learners*

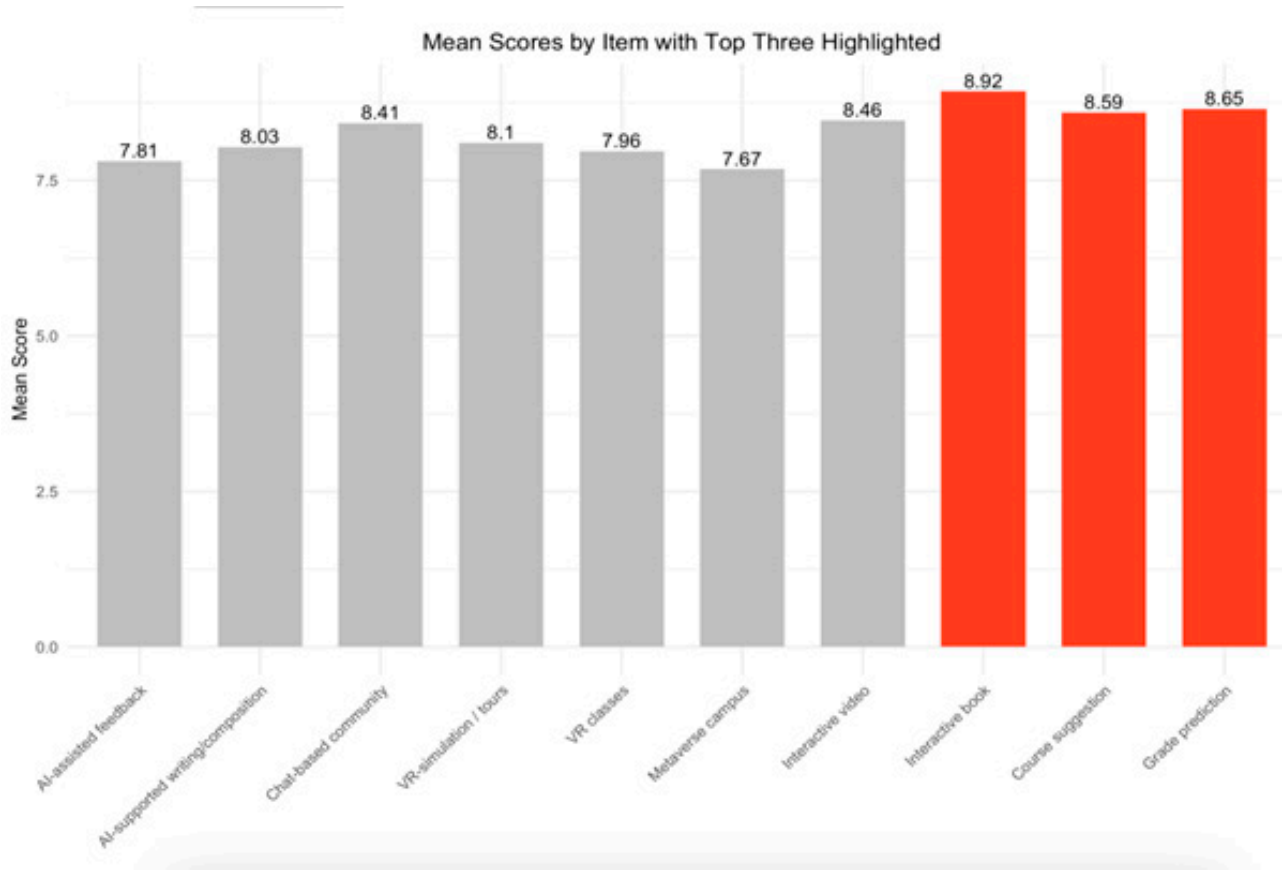

*Comparative Results among Open Universities*

All five open universities in the study had been using the same LMS (Moodle) when the data were collected. Thus, the students had a similar experience and background in terms of technological experience. As illustrated in Table 1, the ratings of students on all items varied across universities.

*Table 1.*
*Means of Perceived Need Ratings for Emerging Technologies*

| | Innovative Pedagogy Technology | UT | STOU | HOU | UPOU | OUM |
|---|---|---|---|---|---|---|
| 1 | Interactive book | 8.94 | 9.00 | 8.44 | 9.57 | 8.68 |
| 2 | Interactive videos | 8.66 | 8.58 | 8.20 | 8.67 | 8.20 |
| 3 | VR simulation/tours | 8.76 | 8.10 | 8.00 | 8.48 | 8.14 |
| 4 | Chat-based community | 8.70 | 8.56 | 8.22 | 8.49 | 8.10 |
| 5 | Course suggestion | 8.96 | 8.94 | 7.70 | 9.30 | 8.04 |
| 6 | VR classes | 8.66 | 8.16 | 7.72 | 7.26 | 8.00 |
| 7 | Grade prediction | 9.14 | 8.96 | 7.90 | 9.27 | 7.98 |
| 8 | Metaverse campus | 8.40 | 7.82 | 7.36 | 7.14 | 7.64 |
| 9 | AI-supported writing/composition | 8.54 | 8.52 | 7.70 | 8.45 | 6.94 |
| 10 | AI assisted feedback | 8.72 | 8.30 | 8.10 | 7.12 | 6.78 |

The series of ANOVA tests comparing responses from students across the five universities showed significant differences on most of the items. For Item 1, results indicated a significant difference, $F(4, 244) = 8.942$, $p < .001$. Similarly, significant differences were found for Item 2, $F(4, 244) = 7.113$, $p < .001$, and for Items 4, 5, 8, 9, and 10, with the following statistics: $F(4, 244) = 2.722$, $p = .030$; $F(4, 244) = 2.962$, $p = .020$; $F(4, 244) = 3.737$, $p = .006$; $F(4, 244) = 7.958$, $p < .001$; and $F(4, 244) = 7.720$, $p < .001$. However, for some items, variations were not found to be statistically significant, as indicated by the results for Item 3, $F(4, 244) = 0.894$, $p = .468$; Item 6, $F(4, 244) = 2.311$, $p = .058$; and Item 7, $F(4, 244) = 0.852$, $p = .493$.

*Figure 4.*
*Boxplots of items with significant variation and CLD labels based on Tukey's HSD*

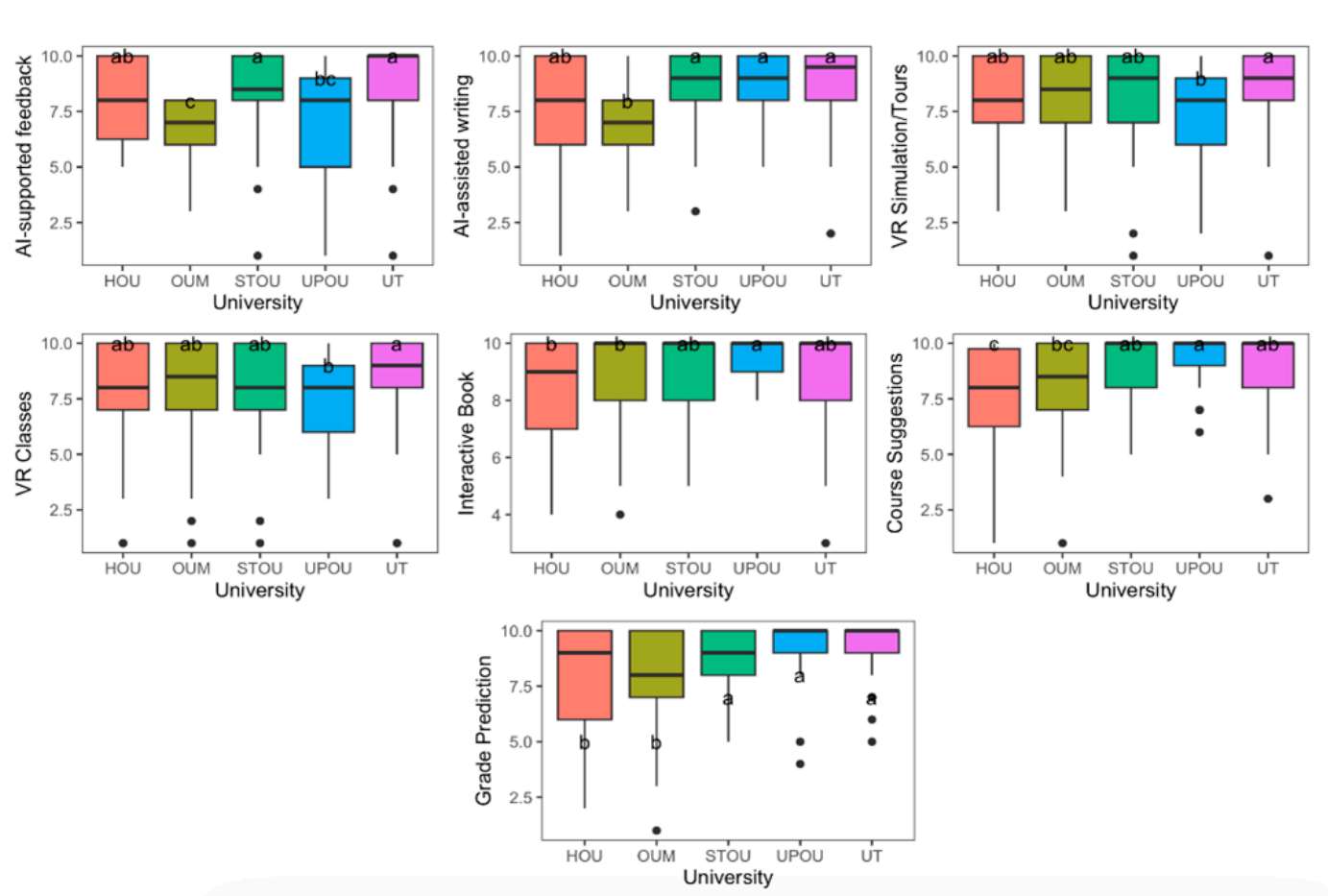

Box plots with compact letter display (CLD) labels are shown in Figure 4 to show the pairwise differences that Tukey's HSD post hoc method found. In the post hoc analysis for Item 1 (AI-assisted feedback), numerous comparisons were statistically significant with 95% family-wise confidence. The mean difference between OUM and HOU was −1.32 (95% CI [−2.38, −0.26], $p = .007$). STOU outscored OUM by 1.52 points (95% CI [0.46, 2.58], $p = .001$). UT had a substantially higher mean score than OUM by 1.94 points (95% CI [0.88, 3.00], $p < .001$). Significantly, UPOU scored 1.18 points lower than STOU (95% CI [−2.25, −0.11], $p = .023$). Finally, UT outscored UPOU by 1.60 points (95% CI [0.53, 2.67], $p = .001$).

In the post hoc analysis for Item 2 (AI-supported writing), many pairwise comparisons between institutions were statistically significant at 95% family-wise confidence. The mean score for STOU was considerably higher than OUM by 1.58 points (95% CI [0.56, 2.60], $p < .001$). UPOU had a considerably higher mean score than OUM, with a difference of 1.51 points (95% CI [0.48, 2.54], $p < .001$). The mean score for UT was substantially higher than OUM, with a difference of 1.60 points (95% CI [0.58, 2.62], $p < .001$).

In the post hoc analysis for Item 4 (VR simulation/tours), UT had a 1.27-point mean score advantage over UPOU (95% CI [0.20, 2.34], $p = .011$. The results were similar in the post hoc analysis for Item 5 (VR classes), where UT had a 1.39-point mean score advantage over UPOU (95% CI [0.23, 2.56], $p = .010$). Other comparisons did not reveal any statistically significant variations in the mean scores for Item 4 across the universities.

In the post hoc analysis for Item 8 (Interactive book), significant differences in mean scores were observed between some university pairs. Notably, the mean score for UPOU was significantly higher than that for HOU, with a mean difference of 1.13 points (95% CI [0.28, 1.98], $p = .003$). Additionally, a significant difference was found between UPOU and OUM, with UPOU scoring 0.89 points higher on average than OUM (95% CI [0.04, 1.74], $p = .035$).

In the post hoc analysis for Item 9 (Course suggestions), multiple pairwise comparisons between universities showed statistically significant mean score differences in the post hoc analysis using Tukey's HSD test for Item 9. Note that STOU's mean score was 1.24 points higher than HOU's (95% CI [0.30, 2.18], $p = .003$). UPOU had a considerably higher mean score than HOU, with a difference of 1.61 points (95% CI [0.66, 2.55], $p < .001$). UT likewise had a 1.26-point mean score advantage against HOU (95% CI [0.32, 2.20], $p = .003$).

In the post hoc analysis for Item 10 (Grade prediction), some pairwise comparisons between universities showed statistically significant mean score differences. STOU's mean score was 1.06 points higher than HOU's (95% CI [0.14, 1.98], $p = .014$). Additionally, UPOU had a substantially higher mean score than HOU, with a difference of 1.37 points (95% CI [0.44, 2.29], $p < .001$). UT likewise had a 1.24-point mean score advantage against HOU (95% CI [0.32, 2.16], $p = .002$). Further comparisons between OUM and other universities showed substantial disparities. STOU, UPOU, and UT had higher mean scores than OUM by 0.98, 1.29, and 1.16 points, respectively.

*Common Underlying Needs*

The results from the three-hour roundtable discussion resulted in identifying the three most common needs across universities that may have been driving the ratings. Figure 5 shows a snapshot of the Mural board containing key themes across the five open universities resulting from processed qualitative data of the survey. The most prevalent needs that emerged were quick or timely feedback, interactivity, engagement, and artificial intelligence-based assistance or support.

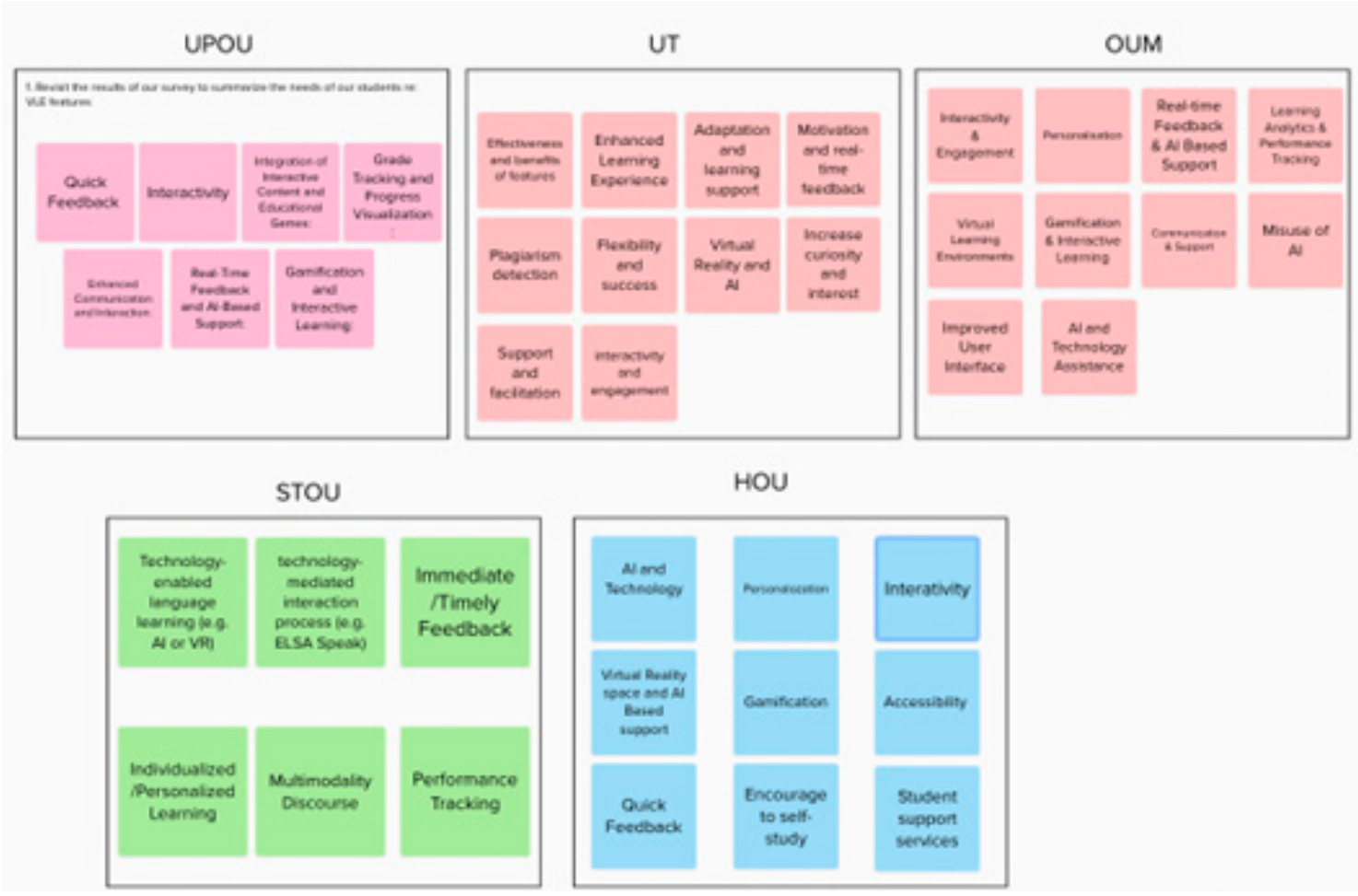

*Figure 5.*
*Mural Board of Perceived Needs among Students of the Five Open Universities*

*Quick or timely feedback*

Timely feedback is important for online learners as it provides them with meaningful learning experiences, which includes allowing students to address their mistakes while improving their academic skills.

A respondent explained that:

> *The ability to make a reply on the feedback given in a specific activity… will help me better improve my academic skills by giving me the freedom to easily reach the checker whenever I need further clarification...*

Additionally, real-time feedback on assignments fosters a sense of progress and helps identify areas for growth:

> *I would be able to learn better from my mistakes.*

One respondent added:

> *I think that providing immediate feedback on our work allows us to identify and correct any mistakes right away. It can also (give us) be our sense of progress, as it tracks our improvement and also identifies our strong and weak areas that need to be worked on.*

These insights underscore the necessity of responsive and interactive feedback systems to enhance student success in the virtual learning environment.

*Interactivity/engagement*

Interactivity and engagement are essential elements of an effective learning environment, especially in virtual settings where encouraging student participation can be challenging. The respondents pointed out features that can promote active collaboration and communication, i.e., games, simulations, and virtual spaces:

> *I hope to experience the educational games and simulations. These will allow students, like me, to put practice into what they are learning. Here they will know their skills and what to do and to avoid. This will also make the course enjoyable.*

> *Since simulation-based learning activities through virtual reality have been mentioned, I would like to suggest making it something like a game of some sort. Nowadays, many are interested in games, which is evident in youth as well as adults. Delivering it in a game-like manner or system can help to interest students in participating.*

Additionally, tools like video lectures, general discussion forums, and alternative learning platforms provide diverse opportunities to enhance interaction and make learning more dynamic and enjoyable:

> *The communication through video calls/ Zoom meetings, we can interact virtually so that we could see each other. It will greatly help us to build communication and friends in our course as well to guide us (in what we do in the course).*

> *Virtual spaces can especially help students feel more ingrained into the academic space and also facilitate meaningful interactions with peers.*

> *...video lectures of professors in charge of a course, put the important points of a topic in focus.*

> *Every course should have a general discussion forum for queries, suggestions, clarifications, and interactions. One's question may also be another student's question.*

These insights emphasize the importance of designing virtual learning environments that are not only accessible but also engaging and interactive to support meaningful educational experiences.

*Artificial intelligence-based assistance or support*

During the roundtable discussion, the respondents highlighted the potential of AI-driven features, such as chatbots for instant support, AI-generated essay checkers for maintaining academic integrity, and voice-activated modules for more natural interaction. These innovations not only simplify tasks like answering queries and managing deadlines but also cater to diverse learning preferences:

> *It would be very helpful if there were chatbots that could respond to students' questions about course requirements and even lectures.*

> *I am still in the process of learning deep technologies, but I would really want to experience a voice-activated module or, like, when you answer, you can just speak what's on your mind. This feature can be compared with Google, Alexa, and Siri. There are times that we cannot type our ideas, and speaking helps to further elaborate it.*

However, alongside these advantages, concerns arose about over-reliance on AI, its impact on authentic learning, and the challenges of keeping pace with evolving technologies, emphasizing the need for balanced and thoughtful integration:

> *... Using emerging technologies can make the learning process easier for many students but at the same time, these emerging techs can put their learning into question because there may be students who are unable to find tools that can keep up with the tech used in their learning especially if that is the only way that the course is taught.*

## DISCUSSION AND CONCLUSION

The research findings indicate a strong demand for interactive and innovative pedagogical approaches supported by emerging technologies in the LMSs of the five open universities. The top three needs identified in the study were the following: the use of interactive books, course suggestions, and grade prediction. Interactive books, according to Sung et al. (2019), can significantly improve students' learning achievements, critical thinking tendencies, and deep motivation. The interactive book effectively boosts students' intrinsic motivation and enables them to achieve successful learning outcomes. It successfully decreases the students' automatic recall of answers, their superficial learning method, and their tendency to focus solely on exams.

On the other hand, course suggestions and grade prediction also play significant roles and can help in decision-making in the educational domain. Predictive analytics in student grades is important in higher education institutions for providing valuable information and developing trusted decision-making, which contributes to data science. In the end, utilizing machine learning to forecast student grades in higher education institutions would improve the decision support system, which would benefit students' future academic performance (Abdul Bujang et al., 2021).

The significant variance found in most of the items reminded us that solutions should be contextualized for each university to ensure that their unique needs are addressed. It was notable to see that UPOU students strongly prefer interactive books compared to students at the other four universities. This may be because many of them have been using these tools and have experienced the benefits of using these technologies over static PDF documents combined with online quizzes and discussion forums. This can be attributed to the feedback received from students regarding their positive experience with the interactive books. A study conducted by Ashrafi et al. (2020) discovered that the level of enjoyment that students report when using an LMS or learning environment influenced their intention to continue using it, highlighting the importance of making learning environments engaging and fun, such as through interactive gamification contents and the use of multimedia formats and graphical interfaces to attract students' attention, which are all offered by interactive books. On the other hand, UT students were prominent in their importance ratings towards VR-based features. The university's strong promotion of VR and metaverse-based initiatives may also help to explain this.

These variations in student preferences reflect the characteristics and institutional strategies of the participating universities. Moreover, the reliance of self-reported data in the survey introduces the possibility of response bias, where students might prioritize features they are already familiar with or perceive as immediately beneficial in a distance learning environment. The use of animated graphics to demonstrate survey items, while helpful for clarity, may have further influenced preferences, as visually intuitive features could have appeared more appealing. These factors underscore the importance of contextualizing findings within the technological and institutional setting of each university.

The results from the roundtable discussion highlight the importance of interactivity, quick feedback, and artificial intelligence in the learning environment. This result is in line with the statement from Pluzhnikova (2020), who said that interactivity during lectures and quick feedback are among the issues raised in teaching in most universities. Including interactive elements in the course material greatly boosts the likelihood that students will find learning enjoyable and that their cognitive abilities will improve (Parker, 2020). These are features that can be found and easily integrated into interactive books, which led us to the conclusion that it may be good to promote their use in online courses for undergraduate students at the five open universities.

The findings related to the use of AI in the learning environment can be supported by Schiff (2022), who stated that AI drastically alters the teacher-student dynamic. The personalized, quick-response AI paradigm is the foundation of AI in education. By resembling teachers in a distance learning environment, AI in education directly addresses one of the major obstacles to distance learning. Nevertheless, it is critical to assess how AI in education will affect educational systems, considering issues with pedagogy, curriculum, international development, teacher automation, ownership of educational decisions, and behavioral manipulation.

Overall, the findings from the study provide guidance on essential 4IR-related technologies that students would most likely embrace and utilize, which contributes to fulfilling the requirements stated by Bozalek and Ng'ambi (2015) for developing graduates with desirable attributes in higher education institutions. Furthermore, the findings from this study would serve as a guide to the participating universities in prioritizing technologies to be featured in the LMSs based on the students' perceived needs. These technologies will not only serve as tools to improve the learning experience of the students but would also contribute to the development of expertise and self-efficacy of learners with these emerging technologies, which could help address their apprehension (Gull et al., 2023) and empower them to face the changing technological landscape of the workforce with confidence. Moreover, the illustrative survey used in the study helped educate the participants regarding emerging 4IR technologies, which could have addressed the gap illustrated by Voogt and Knezek (2018).

The study contributes to ongoing debates on the integration of emerging technologies in education by providing evidence from the context of open and distance elearning institutions in Southeast Asia. As the Fourth Industrial Revolution accelerates technological innovation, there is increasing pressure on educational institutions to equip students with 21st century skills, including critical thinking, adaptability, and digital literacy. By identifying specific technological needs, such as interactive books and AI-driven features, this study emphasizes the importance of aligning technological advancements with student preferences to enhance engagement and learning outcomes. Moreover, the study highlights the critical role of context in shaping the adoption of emerging technologies. While global discourse often focuses on advanced implementations in traditional or hybrid educational settings, this research underscores the challenges and opportunities within ODeL institutions, where accessibility, scalability, and cultural relevance play important roles. By presenting data-driven insights from five leading universities, the study advocates a nuanced approach to technological integration, moving beyond a one-size-fits-all strategy and fostering locally relevant innovation. Additionally, the findings extend current discussions by illustrating the interplay between institutional strategies and student preferences. For instance, the strong preference for virtual reality features at Universitas Terbuka reflects the university's active promotion of VR technologies, demonstrating how institutional initiatives can influence student perceptions. Such insights contribute to the broader debate on how policy and practice shape the effectiveness of educational technologies in diverse settings.

The findings may also help in building a roadmap for technology development and digital transformation among the five open universities. A key component of the roadmap is to prioritize mainstreaming advanced technologies such as learning analytics and interactive learning materials into LMSs. Furthermore, mainstreaming these technologies should focus on enhancing students' engagement, fostering active learning, and providing personalized data-driven learning experiences that cater to the different learning styles and diverse needs of students.

The study's exclusive focus on distance education institutions provides valuable insights into this specific educational context. However, the findings should be interpreted with an awareness of potential biases, including the influence of institutional technological strategies, the reliance of self-reported data, the limitations of voluntary sampling, and the impact of survey design choices. Future research could explore a broader range of distance education institutions to account for regional and cultural differences or incorporate qualitative methods to deepen the understanding of student needs. Such efforts would further enhance the relevance and applicability of the findings for advancing virtual learning environments in the 4IR era.

## ACKNOWLEDGEMENTS

This paper was only possible through the support of the OU5 collaborative research program and the respective research grants of the five open universities in OU5: Hanoi Open University, Open University Malaysia, Sukhothai Thammathirat Open University, Universitas Terbuka, and The University of the Philippines Open University. We would also like to express our gratitude to Mr. Hui Thian Teo and Dr. Christopher Ireland for their invaluable assistance in the design thinking workshop and copyediting of the manuscript.